\documentclass[12pt]{article} 
\usepackage{amsmath}
\usepackage{amssymb}
\usepackage[dvipdfmx]{graphicx}
\usepackage{cite}
\usepackage{comment}
\usepackage{color}
\usepackage[symbol]{footmisc}
\usepackage{here}
\usepackage{siunitx}
\usepackage{lineno}

\newcommand*\patchAmsMathEnvironmentForLineno[1]{
  \expandafter\let\csname old#1\expandafter\endcsname\csname #1\endcsname
  \expandafter\let\csname oldend#1\expandafter\endcsname\csname end#1\endcsname
  \renewenvironment{#1}
     {\linenomath\csname old#1\endcsname}
     {\csname oldend#1\endcsname\endlinenomath}}
\newcommand*\patchBothAmsMathEnvironmentsForLineno[1]{
  \patchAmsMathEnvironmentForLineno{#1}
  \patchAmsMathEnvironmentForLineno{#1*}}
\AtBeginDocument{
\patchBothAmsMathEnvironmentsForLineno{equation}
\patchBothAmsMathEnvironmentsForLineno{align}
\patchBothAmsMathEnvironmentsForLineno{flalign}
\patchBothAmsMathEnvironmentsForLineno{alignat}
\patchBothAmsMathEnvironmentsForLineno{gather}
\patchBothAmsMathEnvironmentsForLineno{multline}
}

\newcommand{\Ecut}{E_{\rm cut}}
\newcommand{\uG}{{\rm \mu G}}

\begin{document}
\makeatletter
\renewcommand{\@cite}[1]{\textsuperscript{#1}} 
\makeatother

\noindent {\Large \bf Potential PeVatron supernova remnant G106.3$+$2.7 seen in the highest-energy gamma rays}\\ \\
\noindent {\large The Tibet AS$\gamma$ Collaboration} \\ 
M.~Amenomori$^1$, 
Y.~W.~Bao$^2$,
X.~J.~Bi$^3$,
D.~Chen$^{4\ast}$,
T.~L.~Chen$^5$,
W.~Y.~Chen$^3$,
Xu~Chen$^{3}$,
Y.~Chen$^2$,
Cirennima$^5$,
S.~W.~Cui$^6$,
Danzengluobu$^5$,
L.~K.~Ding$^3$,
J.~H.~Fang$^{3,7}$,
K.~Fang$^{3}$,
C.~F.~Feng$^8$,
Zhaoyang~Feng$^3$,
Z.~Y.~Feng$^9$,
Qi~Gao$^5$,
Q.~B.~Gou$^3$,
Y.~Q.~Guo$^3$,
Y.~Y.~Guo$^3$,
H.~H.~He$^3$,
Z.~T.~He$^6$,
K.~Hibino$^{10}$,
N.~Hotta$^{11}$,
Haibing~Hu$^{5}$,
H.~B.~Hu$^{3}$,
J.~Huang$^{3}$,
H.~Y.~Jia$^{9}$,
L.~Jiang$^{3}$,
H.~B.~Jin$^{4}$,
K.~Kasahara$^{12}$,
Y.~Katayose$^{13}$,
C.~Kato$^{14}$,
S.~Kato$^{15}$,
K.~Kawata$^{15}$,
W.~Kihara$^{14}$,
Y.~Ko$^{14}$,
M.~Kozai$^{16}$,
Labaciren$^{5}$,
G.~M.~Le$^{17}$,
A.~F.~Li$^{18,8,3}$,
H.~J.~Li$^{5}$,
W.~J.~Li$^{3,9}$,
Y.~H.~Lin$^{3,7}$,
B.~Liu$^{19}$,
C.~Liu$^{3}$,
J.~S.~Liu$^{3}$,
M.~Y.~Liu$^{5}$,
W.~Liu$^{3}$, \\
Y.-Q.~Lou$^{20, 21, 22}$,
H.~Lu$^{3}$,
X.~R.~Meng$^{5}$,
K.~Munakata$^{14}$,
H.~Nakada$^{13}$,
Y.~Nakamura$^{3}$,
H.~Nanjo$^{1}$,
M.~Nishizawa$^{23}$,
M.~Ohnishi$^{15}$,
T.~Ohura$^{13}$,
S.~Ozawa$^{24}$,
X.~L.~Qian$^{25}$,
X.~B.~Qu$^{26}$,
T.~Saito$^{27}$,
M.~Sakata$^{28}$,
T.~K.~Sako$^{15\ast}$,
J.~Shao$^{3,8}$,
M.~Shibata$^{13}$,
A.~Shiomi$^{29}$,
H.~Sugimoto$^{30}$,
W.~Takano$^{10}$,
M.~Takita$^{15}$,
Y.~H.~Tan$^{3}$,
N.~Tateyama$^{10}$,
S.~Torii$^{31}$,
H.~Tsuchiya$^{32}$,
S.~Udo$^{10}$,
H.~Wang$^{3}$,
H.~R.~Wu$^{3}$,
L.~Xue$^{8}$,
Y.~Yamamoto$^{28}$\footnote[2]{deceased},
Z.~Yang$^{3}$,
Y.~Yokoe$^{15}$,
A.~F.~Yuan$^{5}$,
L.~M.~Zhai$^{4}$,
H.~M.~Zhang$^{3}$,
J.~L.~Zhang$^{3}$,
X.~Zhang$^{2}$,
X.~Y.~Zhang$^{8}$,
Y.~Zhang$^{3}$,
Yi~Zhang$^{33}$,
Ying~Zhang$^{3}$,
S.~P.~Zhao$^{3}$,
Zhaxisangzhu$^{5}$
and X.~X.~Zhou$^{9}$ \\ \\
$^1$Department of Physics, Hirosaki University, Hirosaki 036-8561, Japan \\
$^2$School of Astronomy and Space Science, Nanjing University, Nanjing 210093, China \\
$^3$Key Laboratory of Particle Astrophysics, Institute of High Energy Physics, Chinese Academy of Sciences, Beijing 100049, China \\
$^4$National Astronomical Observatories, Chinese Academy of Sciences, Beijing 100012, China \\
$^5$Physics Department of Science School, Tibet University, Lhasa 850000, China \\
$^6$Department of Physics, Hebei Normal University, Shijiazhuang 050016, China \\
$^7$University of Chinese Academy of Sciences, Beijing 100049, China \\
$^8$ Institute of Frontier and Interdisciplinary Science and Key Laboratory of
Particle Physics and Particle Irradiation (MOE), Shandong University, Qingdao 266237, China \\
$^9$Institute of Modern Physics, SouthWest Jiaotong University, Chengdu 610031, China \\
$^{10}$Faculty of Engineering, Kanagawa University, Yokohama 221-8686, Japan \\
$^{11}$Faculty of Education, Utsunomiya University, Utsunomiya 321-8505, Japan \\
$^{12}$Faculty of Systems Engineering, Shibaura Institute of Technology, Omiya 330-8570, Japan \\
$^{13}$Faculty of Engineering, Yokohama National University, Yokohama 240-8501, Japan \\
$^{14}$Department of Physics, Shinshu University, Matsumoto 390-8621, Japan \\
$^{15}$Institute for Cosmic Ray Research, University of Tokyo, Kashiwa 277-8582, Japan \\
$^{16}$Institute of Space and Astronautical Science, Japan Aerospace Exploration Agency (ISAS/JAXA), Sagamihara 252-5210, Japan \\
$^{17}$National Center for Space Weather, China Meteorological Administration, Beijing 100081, China \\
$^{18}$School of Information Science and Engineering, Shandong Agriculture University, Taian 271018, China \\
$^{19}$Department of Astronomy, School of Physical Sciences, University of Science and Technology of China, Hefei, Anhui 230026, China \\
$^{20}$Department of Physics and Tsinghua Centre for Astrophysics (THCA), Tsinghua University, Beijing 100084, China \\
$^{21}$Tsinghua University-National Astronomical Observatories of China (NAOC) Joint Research Center for Astrophysics, Tsinghua University, Beijing 100084, China\\
$^{22}$Department of Astronomy, Tsinghua University, Beijing 100084, China \\
$^{23}$National Institute of Informatics, Tokyo 101-8430, Japan \\
$^{24}$National Institute of Information and Communications Technology, Tokyo 184-8795, Japan \\
$^{25}$Department of Mechanical and Electrical Engineering, Shangdong Management University, Jinan 250357, China \\
$^{26}$College of Science, China University of Petroleum, Qingdao 266555, China \\
$^{27}$Tokyo Metropolitan College of Industrial Technology, Tokyo 116-8523, Japan \\
$^{28}$Department of Physics, Konan University, Kobe 658-8501, Japan \\
$^{29}$College of Industrial Technology, Nihon University, Narashino 275-8575, Japan \\
$^{30}$Shonan Institute of Technology, Fujisawa 251-8511, Japan \\
$^{31}$Research Institute for Science and Engineering, Waseda University, Tokyo 169-8555, Japan \\
$^{32}$Japan Atomic Energy Agency, Tokai-mura 319-1195, Japan \\
$^{33}$Key Laboratory of Dark Matter and Space Astronomy, Purple Mountain Observatory, Chinese Academy of Sciences, Nanjing 210034, China 

\clearpage

\noindent 
{\bf Cosmic rays (protons and other atomic nuclei) are believed to gain energies of petaelectronvolts (PeV) and beyond at astrophysical particle accelerators 
called `PeVatrons' inside our Galaxy.
Although a characteristic feature of a PeVatron is expected to be a hard gamma-ray energy spectrum that extends beyond 100 teraelectronvolts (TeV) without a cutoff,
none of the currently known sources exhibits such a spectrum due to the low maximum energy of accelerated cosmic rays or
insufficient detector sensitivity around 100~TeV.
Here we report the observation of gamma-ray emission from the supernova remnant  
G106.3$+$2.7 [refs. \cite{Joncas1990,Pineault2000}] above 10~TeV. 
This work provides flux data points up to and above 100~TeV and  
indicates that the very-high-energy gamma-ray emission above 10~TeV is well correlated with a molecular cloud\cite{FCRAO} rather than the pulsar
PSR~J2229$+$6114 [refs. \cite{FermiBSL, Fermi2009, Halpern2001a, Hartman1999, MAGIC2010}]. 
Regarding the gamma-ray emission mechanism of G106.3$+$2.7, this morphological 
feature appears to favor a hadronic origin via the 
$\pi^0$ decay caused by accelerated relativistic protons\cite{Naito} over a leptonic one via the inverse-Compton scattering by relativistic electrons\cite{Jones1968,Blum1970}.
Furthermore, we point out 
that an X-ray flux upper limit on the synchrotron spectrum 
would provide important information to firmly establish 
the hadronic scenario as the mechanism of particle acceleration at the source.}

The Milagro experiment reported an elongated gamma-ray source MGRO~J2228$+$61 coincident with PSR~J2229$+$6114 [refs. \cite{Milagro2007, Milagro2009}] at 35~TeV. 
Meanwhile, the VERITAS experiment detected gamma-ray emissions above 1 TeV from the supernova remnant (SNR) 
G106.3$+$2.7 with a flux of $\sim$5\% Crab and named the source
VER~J2227$+$608 [ref. \cite{VERITAS2009}]. 
Recently the HAWC experiment observed G106.3$+$2.7 and reported a best-fit spectrum with an error band above 40~TeV [ref. \cite{HAWC2020}].  
The centroid of VER~J2227+608, 0.4$^\circ$ away from PSR~J2229$+$6114 in the southwest direction, is consistent with that of MGRO~J2228$+$61 and the HAWC centroid within statistical and systematic uncertainties. 
In this work, we use data obtained by the Tibet air shower array combined with the muon detector array (Tibet AS$+$MD) during 719 live days from 2014 February to 2017 May
to observe high-energy gamma-ray emissions from the region around G106.3$+$2.7.
Figure~\ref{2Dmap} shows the detection significance map around G106.3$+$2.7 above 10~TeV, smoothed by the search window size (see Methods).
The events can be well fitted with a symmetrical 2D Gaussian function, and 
the centroid of gamma-ray emissions (a red filled star with a red position error circle) is at (R.A., Dec) = (336.82$^{\circ}$ $\pm$ 0.16$^{\circ}_{stat}$, 60.85$^{\circ}$ $\pm$ 0.10$^{\circ}_{stat}$), 
coincident with the location of a molecular cloud revealed by $^{12}$CO ($J=1-0$) 
emissions (green contours)\cite{FCRAO} overlying the black radio contours\cite{DRAO, CGP} of the SNR, 
and is away from PSR~J2229$+$6114 by 0.44$^{\circ}$ in the southwest direction.
The location of our centroid is also consistent with those of VERITAS\cite{VERITAS2009} and HAWC\cite{HAWC2020}.
Given the distance of 800~pc from the Earth to both PSR~J2229$+$6114 and SNR~G106.3$+$2.7, the distance from the pulsar to the gamma-ray emission centroid obtained by this work is estimated to be 6~pc.
We estimate that our emission centroid deviates from the pulsar location at a confidence level of $3.1\sigma$, based on
the error of $0.14^{\circ}$ including both statistical and systematic errors (see Methods).
Note that the location of the HAWC centroid is consistent with both those of the Boomerang pulsar and the molecular cloud centroid, 
and that the centroids of VERITAS and Fermi are coincident with the location of molecular cloud as well as our centroid.

Figure~\ref{PHI2} shows the distribution of the number of observed incident gamma-ray photons above 10~TeV as a function of the opening angle 
between the estimated arrival direction and the gamma-ray emission centroid.
Fitting the data with a Gaussian function, we estimate the $1\sigma$ extent of the source to be  
$\sigma_{\mathrm{EXT}} = 0.24^{\circ} \pm 0.10^{\circ}_{stat}$, consistent with that 
estimated by VERITAS of 0.27$^{\circ}$ (0.18$^{\circ}$) 
along the major (minor) axis.

Figure~\ref{FLUX} shows the differential energy spectrum of gamma-ray emissions from
G106.3$+$2.7 measured by this work (red filled squares and two red downward arrows for two upper limits). 
The values of our data points can be found in Supplementary Table~1.
The detection significance above 10~TeV is calculated to be 6.1$\sigma$.
This gamma-ray energy spectrum can be fitted by a single power law from 6 to 115~TeV as
$dN / dE = N_0 (E/40~\mathrm{TeV})^{-\Gamma}$ with $N_0 =  (9.5 \pm 1.6_{stat}) \times 10^{-16}$~[cm$^{-2}$~s$^{-1}~$TeV$^{-1}$] and 
$\Gamma = 2.95 \pm 0.17_{stat}$ ($\chi^2$/ndf = 2.5/5), 
and extends above 100~TeV.
The systematic error of $N_0$ is estimated to be $+40\%/-31\%$, resulting from the 12\% uncertainty in the absolute energy scale. 
The flux data points of VERITAS (blue filled circles) are raised by a factor of 1.62 to account for the spill-over of gamma-ray signals outside their window size 
(see Methods).
Our three flux data points below 20~TeV overlapping the energy range covered by the VERITAS flux points 
are statistically consistent with 1.62 times VERITAS's original best-fit power-law function reported in their paper\cite{VERITAS2009}
at the 1.5$\sigma$ level.
Our spectrum is consistent with the HAWC spectrum; 
the $\chi^2$/ndf between the HAWC best-fit power-law spectrum 
and our flux data points overlapping the energy range covered by HAWC 
is 3.0/2, which corresponds to the 0.8$\sigma$ level, when only our statistical errors are considered.
In addition, 
our spectral index above 40~TeV is $\Gamma = 3.17 \pm 0.63_{stat}$, which is consistent with the HAWC index $\Gamma = 2.25 \pm 0.23_{stat}$ at the 1.4$\sigma$ level.

As to the physical mechanism of the gamma-ray emission, both leptonic and hadronic models are possible at the moment; 
the HAWC observation allows for the possibility of a purely leptonic model,
although Bayesian Information Criterion values obtained in fitting their gamma-ray spectrum together with the VERITAS spectrum suggested that a hadronic model is slightly preferred.
We fit the multi-wavelength gamma-ray energy spectrum using the {\it naima} package\cite{naima}, which allows us to estimate the parent particle spectrum so as to best reproduce the observed gamma-ray energy spectrum.
For the energy distribution of the parent particles, we assume an exponential cut-off power-law form of $dN/dE \propto E^{-\alpha} {\rm exp}\left( -E/\Ecut\right)$.
The best-fit gamma-ray spectra for hadronic and leptonic models are shown in Extended Data Figure~1, and the best-fit parameters are listed in 
Supplementary Table~2.
In the hadronic model, 
we get $\Ecut \sim 0.5$~PeV and $\alpha \sim 1.8$. 
The value of $\alpha$ falls 
between that predicted in the standard diffusive shock acceleration ($\alpha = 2$) and 
the asymptotic limit of the very efficient proton acceleration ($\alpha = 1.5$) [refs. \cite{Malkov1999, BE1999}]. 
The total energy of protons with energies $>$1~GeV ($>$0.5~PeV) is estimated to be  
$\sim5.0 \times 10^{47}$~erg ($3.0 \times 10^{46}$~erg) for a target gas density of 10 cm$^{-3}$.
One might argue that, considering the estimated SNR age of 10~kyr, PeV protons escape the SNR much earlier than the 
present time in the standard theory of cosmic-ray acceleration.
Given that $E_{\mathrm{cut}} \sim 0.5$~PeV and that 
the maximum energy of protons remaining inside an SNR is proportional to 
$\tau^{-0.5}$ where $\tau$ is the SNR age\cite{Caprioli2009}, 
protons should be accelerated up to $\sim$1.6 PeV at $\tau = 1$~kyr in the case of G106.3$+$2.7.
This suggests that 
the acceleration of protons at G106.3$+$2.7 should be efficient enough\cite{Caprioli2009} to push their maximum energy up to $\sim$1.6~PeV during 
the SNR free expansion phase. In addition,  G106.3$+$2.7 has a dense molecular cloud nearby indispensable for
accelerated protons to produce TeV gamma rays via $\pi^0$ production.
With $\alpha \sim 1.8$, the proton energy spectrum does not appear softened, 
implying that protons may not be able to escape the SNR easily due to the suppression of the diffusion coefficient (see Supplementary Information).
Future observations of the physical parameters of G106.3$+$2.7 such as the magnetic field and the particle density could provide useful information 
for these theoretical studies on its mechanisms of particle acceleration and confinement. \\
\noindent 
Alternatively, the observed gamma-ray emission might result from protons accelerated by the SNR up to 0.1~PeV and then reaccelerated 
up to 1~PeV by the adiabatic compression of the Boomerang PWN inside the SNR\cite{Ohira2018}.  
If the adiabatic compression ended at an age of 5~kyr as estimated in the paper,
accelerated PeV protons need to travel a distance of 6~pc from the Boomerang PWN to the molecular cloud during the lapse time of $T = 5$~kyr 
until the present time.
The diffusion coeffiicient of a 0.5~PeV proton in the interstellar medium with a magnetic field of $3~\mu$G 
would be $D \sim2\times10^{30}$~cm$^2$/s [ref. \cite{Gabici2009}],
giving a diffusion length of $L \sim 2\sqrt{D T} = 380$~pc [ref. \cite{Atoyan1995}] for $T = 5$~kyr.
Since the diffusion length around an SNR could be shorter by a factor of 10 or more\cite{Fujita2009}, 
we then estimate $L \lesssim 38$~pc. Since this is much larger than 6~pc, it would be possible 
for 0.5~PeV protons to diffuse from the Boomerang PWN 
to the molecular cloud and emit TeV gamma rays through $\pi^0$ production. 
This scenario might not be natural, however, considering that TeV gamma-ray emissions have not been detected from other molecular cloud clumps around the source (see green contours in Figure~1) although protons should also be able to diffuse up to them, and that 
the proton spectrum needs to be kept hard with $\alpha \sim 1.8$ after the diffusion of 6~pc for $T = 5$~kyr.

In the leptonic model, we get $\Ecut \sim 190$~TeV, $\alpha \sim 2.3$ and the SNR magnetic field strength of $\sim$ 9 $\mu$G.
The total energy of relativistic electrons with energies $>10$~MeV is estimated to be 
$\sim1.4 \times 10^{47}$~erg.
We estimate in the Supplementary Information that electrons need to be accelerated freshly within 1~kyr if they originate from the SNR, and that 
electrons provided by the Boomerang PWN are not likely to produce the observed gamma-ray emission in view of the energy budget and the gamma-ray morphology.
The X-ray flux for the small \ang{;2;}-radius region at PSR~J2229$+$6114 has been measured in the 2$\textendash$10~keV range\cite{Halpern2001a}, while
the X-ray flux for the extended region of our gamma-ray emission region with the 1$\sigma$ extent of 0.24$^{\circ}$
has not been published yet, although X-ray data of the region observed by Suzaku, XMM-Newton and Chandra are publicly available\cite{Suzakudata}. 
We point out that a flux upper limit on the synchrotron spectrum at the X-ray band would provide important information to rule out the leptonic scenario 
for particle acceleration at the gamma-ray source (see Supplementary Information Figure~1). 
In a scenario presented in previous papers\cite{Kothes2001, Kothes2006},  
a supernova explosion occurred at or very close to the current location of 
radio pulsar PSR~J2229$+$6114 rather than the center of SNR~G106.3$+$2.7.
Part of the initial shock wave that expanded to the north and east 
encountered a particularly dense H{\scriptsize I} cloud and was quickly decelerated, 
giving rise to a strong asymmetric reverse shock that moved back 
in the southwest direction. Around 6.6~kyr after the supernova explosion
the reverse shock crushed and drove away the initial PWN 
that was forming around the pulsar,
and afterwards a second nebula was formed, which is the current Boomerang PWN of 
age $\sim$3.9~kyr. 
In this scenario, electrons that were contained in the initial PWN might be
blown away by the reverse shock southwestward and somehow reaccelerated at the SNR shell
up to very high energies, emitting gamma rays via inverse-Compton (IC) scattering.
This scenario, therefore, might become possible if the reverse shock velocity
of 6~pc~/~3.9~kyr $\sim$ 1,500~km/s is attainable at the source.
It might also be possible for unknown nearby pulsars to contribute to the observed gamma-ray emission.

In addition, a hybrid (leptonic$+$hadronic) scenario might also be possible\cite{Bartko2008}. 
If the birthplace of the pulsar was coincident with the location of the molecular cloud,
both electrons and protons accelerated during the early age of the pulsar 
could contribute to the observed TeV gamma-ray emission via IC scattering and $\pi^0$ decay ($p + p \rightarrow \pi^0 \rightarrow 2\gamma$), respectively.
This scenario might not be natural, however, since no enhancement of gamma-ray emission was observed by {\it Fermi} at the current location of the pulsar\cite{Fermi2019}
although VERITAS detected some gamma-ray excess events there\cite{VERITAS2009}.
If this scenario applies, the pulsar moved $\sim$0.4$^{\circ}$ 
towards its current location with a transverse velocity of $\sim$570 km/s during the age of 10~kyr.
Future measurements of the pulsar velocity would be important to investigate the validity of this scenario.

It is known that the energy spectrum of hadronically-induced gamma rays rises steeply below $\sim$200~MeV and approximately follows the energy spectrum 
of parent particles above a few GeV, resulting in a characteristic ``$\pi^0$-decay bump'' in the gamma-ray spectrum. 
Hopefully, further multi-wavelength observations in the future would establish the hadronic origin of gamma-ray emissions from SNR G106.3$+$2.7.

{}

\section*{Correspondence}
Correspondence and requests for materials should be addressed to 
T.~K.~S. (tsako@icrr.u-tokyo.ac.jp), D.~C. (chending@nao.cas.cn), 
J.~H. (huangjing@ihep.ac.cn), M.~O. (ohnishi@icrr.u-tokyo.ac.jp),
M.~T. (takita@icrr.u-tokyo.ac.jp) and X.~Z. (xiaozhang@nju.edu.cn).

\section*{Acknowledgments}
The collaborative experiment of the Tibet Air Shower Arrays has been
conducted under the auspices of the Ministry of Science and Technology
of China and the Ministry of Foreign Affairs of Japan. 
This work was supported in part by a Grant-in-Aid for Scientific Research on Priority Areas from the Ministry of Education, Culture, Sports, Science and Technology, 
and by Grants-in-Aid for Science Research from the Japan Society for the Promotion of Science in Japan.
This work is supported by the National Key R\&D Program of China (No. 2016YFE0125500),
the Grants from the National Natural Science Foundation of China (Nos. 11533007, 11673041, 11873065, 11773019, 11773014, 11633007, 11803011, 
and 11851305), and the Key Laboratory of Particle Astrophysics, Institute of High Energy Physics, CAS.
The research presented in this paper has used data supplied through the Canadian Galactic Plane Survey.
This work is also supported by the joint research program of the Institute for Cosmic Ray Research (ICRR), the University of Tokyo.

\section*{Author contributions}
The whole Tibet AS$\gamma$ collaboration contributed to the publication 
in terms of various aspects of the research ranging from hardware-related issues
such as the design, construction, maintenance, calibration, etc. of the instrument to 
software-related issues such as data reduction, data analysis, MC simulation, astrophysical explanation, etc.
D.~C., J.~H., M.~O., T.~K.~S., M.~T. and X. Z. analyzed the data and prepared the manuscript. 
All the authors discussed the results of this work and commented on the manuscript.
Authors to whom correspondence should be addressed: 
T.~K.~S. (tsako@icrr.u-tokyo.ac.jp), D.~C. (chending@nao.cas.cn), 
J.~H. (huangjing@ihep.ac.cn), M.~O. (ohnishi@icrr.u-tokyo.ac.jp),
M.~T. (takita@icrr.u-tokyo.ac.jp) and X.~Z. (xiaozhang@nju.edu.cn).

\section*{Competing Interests}
The authors declare no competing financial interests. 

\clearpage
\begin{figure}[H]
  \begin{center}
  \includegraphics[width=0.7\textwidth]{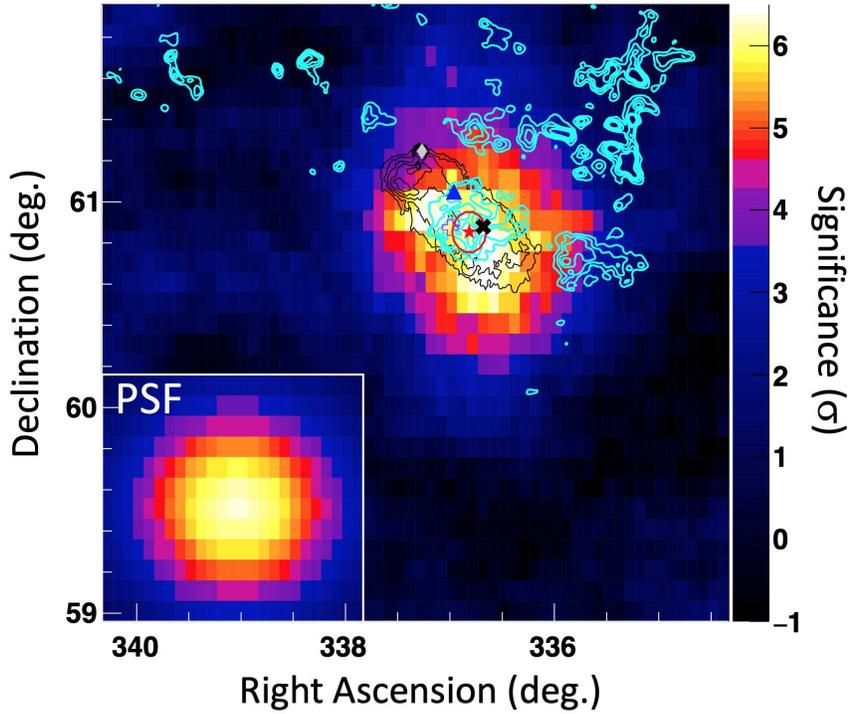}
  \end{center}
  \caption{{\bf Significance map around SNR G106.3$+$2.7 as observed by Tibet AS$+$MD above 10~TeV.} 
The inset figure shows our point spread function (PSF).
The red filled star with a 1$\sigma$ statistical position error circle is the centroid of gamma-ray emissions determined by this work, 
while the magenta open cross, the black X mark and the blue filled triangle are the centroids determined by VERITAS\cite{VERITAS2009}, {\it Fermi}\cite{Fermi2019} and HAWC\cite{HAWC2020}.
The black contours indicate 1420~MHz radio emissions from the Dominion Radio Astrophysical Observatory Synthesis Telescope\cite{DRAO, CGP}, and
the cyan contours indicate $^{12}$CO emissions from the Five College Radio Astronomy Observatory survey\cite{FCRAO}.
The gray filled diamond at the northeast corner of the radio emission marks the pulsar PSR~J2229$+$6114.}
  \label{2Dmap}
\end{figure}
\begin{figure}[H]
  \begin{center}
  \includegraphics[width=0.7\textwidth]{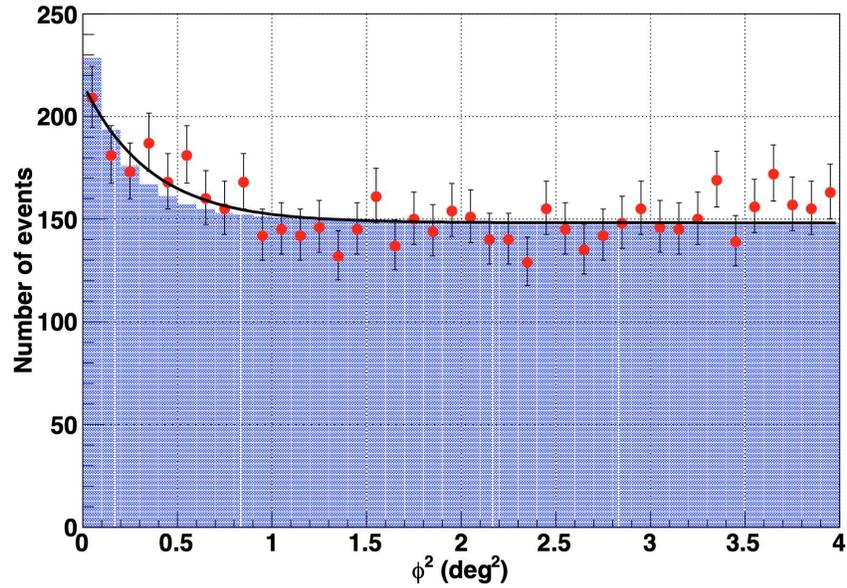}
  \end{center}
\caption{{\bf Projected angular distribution of events observed above 10~TeV.}
The horizontal axis $\phi^2$ represents the square of the opening angle between the estimated event arrival direction and the centroid of gamma-ray emissions
  determined by this work. The red filled circles with 1$\sigma$ statistical error bars are the experimental data with the best-fit black solid curve (see Methods). The blue histogram is 
the expected event distribution by our MC simulation assuming a point-like gamma-ray source. }
  \label{PHI2}
\end{figure}
\begin{figure}[H]
  \begin{center}
  \includegraphics[width=0.7\textwidth]{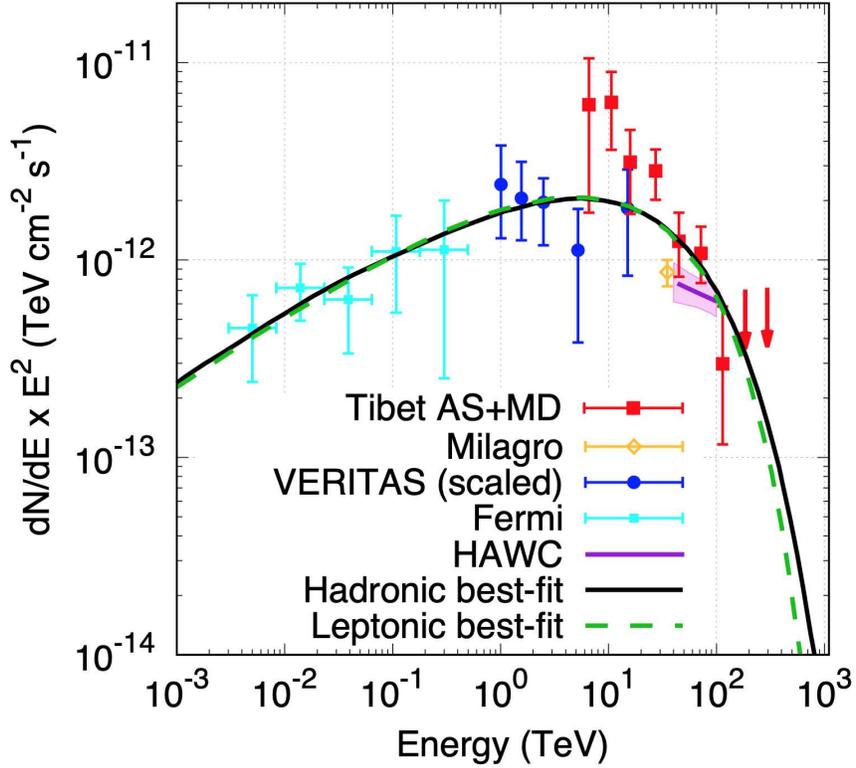}
  \end{center}
 \caption{{\bf Differential energy spectrum of gamma-ray emissions from SNR G106.3$+$2.7.}
Red filled squares (Tibet AS$+$MD) represent data measured by this work 
with two 99\% C.L. upper limits (downward red arrows), VERITAS\cite{VERITAS2009} (deep-blue filled circles), {\it Fermi}\cite{Fermi2019} (sky-blue crosses), 
Milagro\cite{Milagro2009} (an orange open diamond) and HAWC\cite{HAWC2020} (a purple solid line with a shaded light purple area indicating the 1$\sigma$ statistical error band).
The error bars represent the 1$\sigma$ statistical uncertainty.
VERITAS's data points are raised by a factor of 1.62 from the original values (see Methods).
The black solid (green dashed) line is the best-fit curve of the hadronic (leptonic) model for the combined data points of Tibet AS$+$MD, VERITAS and {\it Fermi}. 
}
  \label{FLUX}
\end{figure}

\clearpage

\section*{Methods}
\subsection*{Experiment}
The Tibet air-shower (AS) array has been in operation since 1990 at Yangbajing (90.522$^{\circ}$E, 30.102$^{\circ}$N; 4300 m above sea level) in Tibet, China, to observe cosmic rays and gamma rays above TeV energies\cite{Tibet1992}. 
Currently the AS array covers an area of $\sim$65,700~m$^2$ with 597 plastic scintillation counters placed on a 7.5-m square grid.
An event trigger signal is issued on condition that any four-fold coincidence occur among the counters recording more than 0.6 particles each. 
The muon detector (MD) array has been in operation since 2014, covering an area of 3,450~m$^2$ with 64 water cells constructed approximately 2.4~m 
under the surface AS array.  
Air-shower muons with energies $\gtrsim$~1~GeV penetrate the soil overburden ($\sim$19 radiation lengths), while the soil shields electromagnetic components 
($e^{\pm}$ and $\gamma$) in air showers.
Each water cell is filled with clear water with a depth of 1.5~m and its inner walls are covered with white Tyvek sheets. 
A 20-inch downward-facing photomultiplier tube put on the ceiling of each cell detects Cherenkov photons emitted by penetrating air-shower muons and reflected by the inner walls.
Essentially, the MD array measures the number of muons in air showers that have triggered the AS array.
The MD array records charge and timing information from the water cells in synchronization with event trigger signals issued by the AS array. 
Using the MD array, we can dramatically reduce background cosmic-ray events in gamma-ray observation by selecting 
muon-poor air-shower events, 
since air showers induced by background cosmic rays contain far more muons than those induced by primary gamma rays.
Details are provided in our recent paper\cite{Tibet2019} reporting the detection of cosmic gamma rays beyond 100~TeV.

\subsection*{MC simulation}
Air showers are generated along the orbit of SNR~G106.3$+$2.7 within a zenith-angle range of $\theta < 60^{\circ}$, 
assuming a gamma-ray energy spectrum with a power-law index of $-2.9$ above 0.3~TeV. 
CORSIKA v7.4000 [ref. \cite{corsika1998}] is used for air-shower generation, with EPOS-LHC\cite{corsika2015} for the high-energy hadronic interaction model and 
FLUKA v2011.2b [refs. \cite{fluka2005, fluka2014}] for the low-energy hadronic interaction model. 
The generated air-shower particles are fed into the detector response simulation of the AS array and the MD array developed by GEANT v4.10.00 [ref. \cite{geant2003}].
Detailed simulation procedures can be found in our previous papers\cite{Sako2009, Tibet2019}.

\subsection*{Data analysis}
We estimate the arrival direction of a primary particle based on the relative timing information of the AS counters assuming a cone-shaped air-shower front. 
The slope of the cone is optimized by the MC simulation for gamma-ray observation.
The angular resolution (50\% containment) is estimated to be 0.5$^{\circ}$ and 0.2$^{\circ}$ for 10~TeV and 100~TeV gamma rays, respectively.
The energy of a primary particle is reconstructed from the detected particle densities of the AS counters.
Above 10~TeV, the lateral distribution of particle densities is fitted by 
the Nishimura-Kamata-Greisen (NKG) function, and then the particle density 
50~m away from the air-shower axis ($S$50) is calculated from the best-fit NKG function.
The reconstructed energy of the primary photon $E_{\mathrm{rec}}$ is thus obtained as a function of $S$50 and the zenith angle.
The energy resolution is estimated to be 40\% at 10~TeV and 20\% at 100~TeV for primary gamma rays\cite{kawata2017}.
The purity of $E_{\mathrm{rec}}$ bins for $10\textendash16$ TeV, $40\textendash63$ TeV and $100\textendash158$ TeV is
34\%, 45\% and 55\%, respectively, while
the contamination from lower (higher) energies is 39\% (27\%), 34\% (21\%) and 30\% (15\%), respectively.
Below 10~TeV, the energy of a primary photon is reconstructed from 
$\Sigma \rho$, which is the sum of detected particle densities of all AS counters.
The uncertainty in the absolute energy scale is estimated to be 12\% [ref. \cite{Moon}].
We evaluate our pointing precision at the declination of G106.3$+$2.7 by re-analyzing the source location of the Crab Nebula.
After thinning out air-shower events so that the zenith-angle distribution of events is adjusted to that of G106.3$+$2.7, 
we fit the events in the same way described below. As a result, we obtain Crab's position as
(R.A., Dec.) = ($83.636^{\circ} \pm 0.137^{\circ}_{stat}$, $21.991^{\circ} \pm 0.099^{\circ}_{stat}$), and  
the deviation from the location of the Crab pulsar as $0.003^{\circ} \pm 0.137^{\circ}_{stat}$ in R.A. and $0.024^{\circ} \pm 0.099^{\circ}_{stat}$ in Dec.
Therefore, we estimate the systematic pointing error for G106.3$+$2.7 as $0.10^{\circ}$ in angular distance, although 
we expect that, as we accumulate statistics, the pointing error would be reduced to $0.023^{\circ}$ 
(from our observation of the Crab Nebula\cite{Tibet2019}), and further down to
$< 0.011^{\circ}$ (from the analysis of the cosmic-ray shadow of the Moon\cite{Moon}).

The single peak of each MD cell is defined as the peak of the charge distribution of air-shower events that have triggered the AS array. 
The number of muons $N_{\mu}$ is calculated for each MD cell by dividing the recorded charge by the single peak, and then the total sum $\Sigma N_{\mu}$ is obtained 
for each air-shower event by summing up the $N_{\mu}$ values from all the MD cells.

The event selection criteria are the same as in our previous work\cite{Tibet2019}, except
that we re-optimize the $\Sigma N_{\mu}$ condition, 
namely, $\Sigma N_{\mu} < 0.15 \left(\Sigma \rho \right)^{0.28}$ or $\Sigma N_{\mu} < 3.6\times 10^{-4} \left( \Sigma \rho \right)^{1.4}$.

We adopt the Equi-Zenith-Angle method employed in our previous work\cite{Moon, Tibet2015} to estimate background and gamma-ray excess counts. 
In this work, we take 20 off-sources with the same size and zenith angle as the on-source.
The radius of the search window is optimized
as $R_{\mathrm{w}} = 6.9^{\circ} / \sqrt{\Sigma \rho}$, with a lower limit of 0.5$^{\circ}$ to keep $>$90\% gamma-ray events at high energies $\gtrsim$ 100~TeV where background contamination is low. 

In Figure \ref{2Dmap}, the celestial region around SNR~G106.3$+$2.7 is gridded in $0.1^{\circ} \times 0.1^{\circ}$ pixels.
The significance value of each pixel is calculated\cite{LiMa} from background and gamma-ray excess counts within a search window centered at the pixel 
with the variable radius of $R_{\mathrm{w}}$.
To derive the centroid of gamma-ray emissions observed above 10~TeV, 
we fit the events within the $5^{\circ} \times 5^{\circ}$ region 
around the SNR using the unbinned maximum likelihood method.

To derive the 1$\sigma$ extent of the source $\sigma_{\mathrm{EXT}}$, we fit the data in Figure \ref{PHI2} with a Gaussian function: 
\begin{equation}
G(\phi^2; A, \sigma_{\mathrm{EXT}}) = A \exp\left[ - \dfrac{\phi^2}{2(\sigma_{\mathrm{PSF}}^2 + \sigma_{\mathrm{EXT}}^2)}  \right] + N_{\mathrm{BG}}, 
\end{equation}
where $A$ and $\sigma_{\mathrm{EXT}}$ are two fitting parameters, while
the number of background events $N_{\mathrm{BG}} = 148$ and the PSF of the instrument 
$\sigma_{\mathrm{PSF}} = 0.35^{\circ}$ are estimated 
from the background cosmic-ray data and the gamma-ray MC simulation, respectively. 

In Figure~\ref{FLUX}, VERITAS's flux data points are raised by a factor of 1.62 from the original values\cite{VERITAS2009}.
Using the source extension reported by VERITAS with the 1$\sigma$ angular extent of $0.27^{\circ}$ ($0.18^{\circ}$) along the major (minor) axis and their PSF of 0.11$^{\circ}$, we estimate the spill-over of gamma-ray signals outside their integration region of radius 0.32$^{\circ}$ to be 38.3\%, which means that
the total flux of the source should be higher by a factor of $1/(1-0.383) = 1.62$.

\subsection*{Data Availability} 
The data that support the plots within this paper and other findings of this study are available from the website of the Tibet AS$\gamma$ Collaboration
(https://www.tibet-asg.org) or from the corresponding authors upon reasonable request.

\subsection*{Code Availability} 
The codes used in this work are embedded within the analysis framework of the Tibet AS and MD array, and it is not practically possible to extract them.
The codes, therefore, are not publicly available. 

{}


\clearpage

\begin{center}
\fbox{\Large Supplementary Information}
\end{center}
\section{Flux data points of the gamma-ray energy spectrum}
The following table shows the flux data points measured by this work along with the detection significance values\cite{LiMa}.
\begingroup
\renewcommand{\arraystretch}{1.3}
\begin{table}[H]
  \caption{{\bf Photon flux data points and detection significances for SNR~G106.3$+$2.7.}}
\vspace{-5mm}
  \label{sig}
  \begin{center}
    \begin{tabular}{ccc}
      \hline
Energy (TeV) & Flux (TeV$^{-1}$ cm$^{-2}$ s$^{-1}$) & Significance ($\sigma$) \\
      \hline
6.6 &   $(1.4 \pm 1.0) \times 10^{-13}$ & 1.9\\
11 & $(5.6 \pm 2.4) \times 10^{-14}$ & 2.5\\
16 & $(12.5 \pm 5.7) \times 10^{-15}$ & 2.3 \\
27 & $(3.8 \pm 1.1) \times 10^{-15}$ & 3.9\\
45 & $(6.2_{-2.1}^{+2.4} ) \times 10^{-16}$ & 3.6\\
72 & $(21.0_{-6.2}^{+7.7} ) \times 10^{-17}$ & 4.9\\
114 & $(2.3_{-1.4}^{+2.2} ) \times 10^{-17}$ & 2.2\\
184 &  $2.1 \times 10^{-17}$ (99\% U.L.) &  \textendash\\
295 & $8.3 \times 10^{-18}$ (99\% U.L.) &  \textendash\\
      \hline
    \end{tabular}
  \end{center}
\end{table}
\endgroup

\section{Modelling of the multi-wavelength spectrum}
\indent We investigate the energy distribution of parent particles (electrons or protons) for gamma-ray emissions from the supernova remnant (SNR) G106.3$+$2.7 using the {\it naima} package\cite{naima}, which
allows us to perform the Markov chain Monte Carlo fitting of radiative models to the observed 
gamma-ray energy spectrum.
We use the data measured by {\it Fermi}\cite{Fermi2019}, VERITAS\cite{VERITAS2009} and this work in the fitting of the hadronic model.
The VERITAS data points are raised by a factor of 1.62 to account for the spill-over of 
gamma-ray signals outside their window size of 0.32$^{\circ}$ radius (see Methods of the letter).
In the fitting of the leptonic model, we also include the data provided by the Dominion Radio Astrophysical Observatory's Synthesis Telescope\cite{Pineault2000}.
We assume the distribution of parent particles to have an exponential cutoff power-law form of
$dN/dE = A E^{-\alpha} {\rm exp}\left( -E/\Ecut\right)$ where $A$, $\alpha$ and $\Ecut$ are
three fitting parameters.
The normalization factor $A$ is replaced by the total energy of parent particles $W_{e/p}$ in the fitting procedure.

\subsection{The leptonic scenario}
\indent Taking into account the Cosmic Microwave Background (CMB) photons and 
infrared (IR) photons\cite{Porter2006}
with a temperature of 30~K and an energy density of $1\ {\rm eV\ cm}^{-3}$, we attribute the observed gamma-ray energy spectrum to the inverse Compton (IC) scattering of these very low energy photons by relativistic electrons accelerated at the SNR.
The best-fit parameters of the parent electron energy distribution are listed in Table~\ref{tab:para}, and the corresponding gamma-ray energy spectrum is plotted 
in Figure~\ref{fig:lep}(a).
From the fitting results, $\alpha$ and $\Ecut$ are estimated to be 2.3 and 190~TeV, respectively. 
The synchrotron cooling time of relativistic electrons is given by 
$\tau_{syn}\approx500\ (E/1\ {\rm PeV})^{-1}(B/5\ \uG)^{-2}$~yr. 
For a magnetic field strength $B \sim 8.6~\mu$G, the 190~TeV electrons have $\tau_{syn} \approx 0.9$~kyr, which is much shorter than the SNR age of about 10~kyr.
This means that these electrons should be accelerated freshly within a thousand years if they are injected from the SNR.

\indent Another possible source of these electrons is the Boomerang pulsar PSR~J2229$+$6114 and its pulsar wind nebula (PWN), which is $0.44^{\circ}$ (or 6 pc at a distance of 0.8 kpc) away from the centroid of observed gamma-ray emissions.
The electrons accelerated at the termination shock of this pulsar might be able to diffuse into the 
gamma-ray emission region.
The required total energy of electrons is $\sim1.4\times10^{47}$~erg, which only takes up 
$\sim2\%$ of the spin-down energy released in the entire pulsar lifetime.
If the rest of the spin-down energy goes into the magnetic field, the average magnetic field in the PWN would be much larger than the required value of 8~$\mu$G and results in very large 
fluxes at radio and X-ray wavelengths.
The energy budget may be reconciled if the true age of PSR~J2229$+$6114 is 
much shorter than its characteristic age, say $\sim$1~kyr. 
In this case, however, it would be difficult to explain the spatial separation
between PSR~J2229$+$6114 and the observed $\gamma$-ray emission region.
The diffusion length is given by $R_d=2\sqrt{D(E)T}$ [ref. \cite{Aharonian1996}], where $D(E)$ is 
the energy-dependent diffusion coefficient that can be
expressed as $D(E)=\chi 10^{28} (E/10\ {\rm GeV})^{0.5} (B / 3\ \mu{\rm G})^{-0.5}~{\rm cm^2\ s^{-1}}$ [ref. \cite{Gabici2009}] with a suppression factor $\chi$.
Here we simply assume the energy dependence of the average Galactic diffusion coefficient ($D \propto E^{0.5}$), although its validity is not 
obvious around pulsars.
To produce the offset of $0.44^{\circ}$ or 6 pc for 190~TeV electrons within their lifetime 
of $\tau_{syn} \approx 0.9$~kyr, the suppression factor $\chi$ should be $\sim$0.004.
With these values, 1~TeV electrons that mainly generate $\sim$10~GeV photons via the IC scattering process diffuse away only by 1.7 pc or $\sim 0.12^{\circ}$ during the age 
of 1~kyr.
This implies that the GeV gamma-ray emission should be spatially coincident with 
PSR~J2229$+$6114 rather than the TeV gamma-ray emission, which is inconsistent with 
the morphology reported by {\it Fermi}.

\indent 
In Figure~\ref{fig:lep}(a), the flux of PSR~J2229$+$6114 in the 2$\textendash$10~keV range\cite{Halpern2001a} is indicated by the gray open diamond, while
an X-ray flux upper limit for the observed gamma-ray emission region has not been measured so far.
We emphasize that an upper limit on the synchrotron flux at the X-ray band would provide crucial information to rule out the leptonic scenario 
as the mechanism of particle acceleration at the source. 


\subsection{The hadronic scenario}
\begin{figure}[H]
\includegraphics[width=1.0\textwidth]{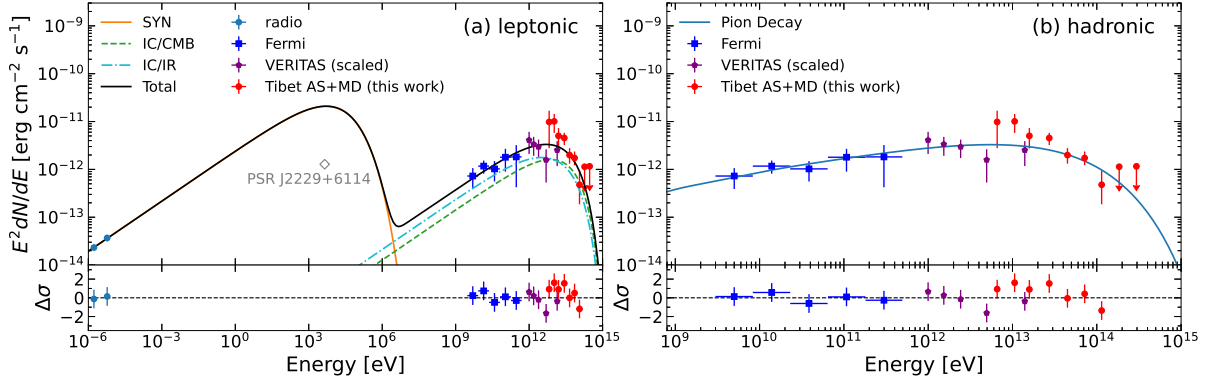}
\caption{{\bf Spectral gamma-ray energy distribution of G106.3+2.7.}
(a) The flux data points with 1$\sigma$ statistical error bars include measurements by Tibet AS$+$MD (red dots; this work), 
{\it Fermi}\cite{Fermi2019} (blue squares), VERITAS\cite{VERITAS2009} (purple pentagons) and 
the Dominion Radio Astrophysical Observatory's Synthesis Telescope\cite{Pineault2000} (turquoise blue dots). 
The two red downward arrows above $10^{14}$~eV show 99\% C.L. upper limits obtained by this work.
Note that all the VERITAS data points are raised by a factor of 1.62 to account for the spill-over of gamma-ray signals outside their window size of 0.32$^{\circ}$ radius.
The best-fit gamma-ray energy spectrum in the leptonic model is shown by the black solid curve, with the flux by the electron synchrotron radiation (the orange solid curve), 
the IC scattering of CMB photons (the green dashed curve) and 
the IC scattering of IR photons (the light blue dash-dotted curve). 
The gray open diamond shows the flux of PSR~J2229$+$6114 obtained in the 2$\textendash$10~keV range\cite{Halpern2001a}. 
(b) The best-fit gamma-ray energy spectrum in the hadronic model is shown by the turquoise blue solid curve.
The lower panels show the residual $\Delta \sigma$ of the fit.
}
\label{fig:lep}
\end{figure}

\indent In the hadronic model, $\gamma$-rays are produced via proton-proton inelastic 
collisions to produce $\pi^0$ particles which subsequently decay into energetic photons. 
We assume the proton density of the target gas\cite{Fermi2019} 
to be $n_{\rm t}=10\ {\rm cm}^{-3}$.
The best-fit parameters of the parent proton energy distribution are listed in Table~\ref{tab:para}, and the corresponding gamma-ray energy spectrum is displayed
in Figure~\ref{fig:lep}(b).
With $\Ecut \sim$ 500~TeV, the proton spectrum has a power-law index of
$\alpha \sim 1.8$, falling 
between the index of the standard diffusive shock acceleration ($\alpha = 2$) and 
the asymptotic limit
of the very efficient proton acceleration\cite{Malkov1999, BE1999} ($\alpha = 1.5$).
The most likely source of these protons is SNR~G106.3+2.7.
Generally, SNRs are expected to accelerate protons up to very high energies 
during their early stage, and the accelerated protons may escape and diffuse away from their acceleration site, resulting in a softened proton energy spectrum.
With $\alpha \sim 1.8$, however, the proton energy spectrum does not seem to be
softened, which implies that the diffusion may not play a noticeable role.
Very likely the protons cannot escape the SNR easily due to the suppression of the diffusion coefficient\cite{Fujita2010}; 
while the diffusion length of 500~TeV protons would be 50~pc 
for $\chi=0.01$ and $B = 3~\mu$G, protons can be trapped inside a smaller volume
comparable to the size of the gamma-ray emission region 
if $\chi<0.01$ and/or the true SNR age is shorter than 10~kyr.
Alternatively, 500~TeV protons 
can be confined in an even smaller region of $\sim$5~pc in radius with a magnetic field of 100~$\uG$ in the extreme case of the Bohm diffusion. 
The spatially-integrated spectrum of protons trapped inside a small volume 
would not be modified by the diffusion process.

\clearpage

\begingroup
\renewcommand{\arraystretch}{1.3}
\begin{table}[H]
\centering
\caption{{\bf Best-fit parameters of the energy distribution of parent particles.}
Electrons (protons) are assumed for the parent particles of the leptonic (hadronic) model.\label{tab:para}}
\begin{tabular}{cccccc}
\hline
             & $\alpha$ & $\Ecut$ (TeV) & $W_{e/p}^{\ddag}$ (10$^{47}$ erg) & $B$ ($\mu$G) & $\chi^2$/ndf\\
\hline
leptonic & $2.30^{+0.08}_{-0.07}$ & $190^{+127}_{-66}$ & $1.4^{+1.8}_{-0.7}$ & $8.6^{+3.4}_{-2.5}$ & 12.8/15\\
\medskip 
hadronic& $1.79^{+0.08}_{-0.09}$ & $499^{+382}_{-180}$ & $5.0^{+0.7}_{-0.6}$ & --- & 13.0/14\\
\hline
\end{tabular}
\\
{\raggedright $^{\ddag}$$W_{e/p}$ is the total energy above 10 MeV and 1 GeV for electrons and protons, respectively, where $W_p$ is for a target gas density of 10~cm$^{-3}$.  \par}
\end{table}
\endgroup

\clearpage

\end{document}